\documentclass[lettersize,journal]{IEEEtran}

\usepackage{subcaption}
\usepackage{graphicx}
\usepackage{float} 

\usepackage{amsmath,amsfonts}
\usepackage{algorithmic}
\usepackage{algorithm}
\usepackage{array}
\usepackage[caption=false,font=normalsize,labelfont=sf,textfont=sf]{subfig}
\usepackage{textcomp}
\usepackage{stfloats}
\usepackage{url}
\usepackage{verbatim}
\usepackage{graphicx}
\usepackage{cite}

\usepackage{xcolor}

\begin{document}





\title{Quantum Machine Learning-based 6G Network: Enabling Adaptive Communication and Model Aggregation}

\author{Wenjing Xiao, Jiatai Yan, Chenglong Shi, Shixin Chen, Miaojiang Chen, Min Chen,~\IEEEmembership{IEEE, Fellow}, \\ Saif~Al-Kuwari, Ahmed Farouk
\thanks{W. Xiao, J. Yan, C. Shi, S. Chen and M. Chen are with the School of Computer, Electronics and Information, Guangxi University, Nanning 530004, China, and also with the Guangxi Key Laboratory of Multimedia Communications and Network Technology, Guangxi 530004, China, E-mail: wenjingx@gxu.edu.cn, jiataiyan@st.gxu.edu.cn, chenglongs@st.gxu.edu.cn, sxchen@st.gxu.edu.cn, mjchen\_cs@gxu.edu.cn.}
\thanks{M. Chen is with the School of Computer Science and Engineering, South China University of Technology, Guangzhou 510006, China, and also with the Pazhou Laboratory, Guangzhou 510330, China, E-mail: minchen@ieee.org.}
\thanks{Saif~Al-Kuwari is with the Qatar Center for Quantum Computing, College of Science and Engineering, Hamad Bin Khalifa University, Doha, Qatar. E-mail: smalkuwari@hbku.edu.qa.}
\thanks{A. Farouk is with School of of Faculty of Computers and Artificial
Intelligence, Hurghada University, Hurghada, Egypt, E-mail: ahmed.farouk@sci.svu.edu.eg.}
}

\markboth{}%
{Shell \MakeLowercase{\textit{et al.}}: A Sample Article Using IEEEtran.cls for IEEE Journals}

\maketitle

\begin{abstract}
With the advent of sixth-generation (6G) mobile communication technology, vehicle-to-everything (V2X) communication faces unprecedented challenges in communication efficiency, system generalization capabilities, and model collaboration. 
Conventional machine learning struggles with high-dimensional state spaces, slow convergence, and poor generalization under heterogeneous V2X nodes, rapidly varying channels, and multimodal sensing data in V2X systems.
To address these issues, we propose a quantum-enhanced framework for V2X communication and model aggregation that targets efficient, robust, and intelligent transportation in 6G, which includes four modules: the channel-adaptive semantic communication module, the multimodal fusion module, the model transfer module, and the federated aggregation module. Specifically, the channel-adaptive semantic communication module leverages quantum convolutional neural networks (CNN) and quantum distortion metrics to enable efficient transmission and strong generalization across diverse conditions. The multimodal fusion module exploits quantum attention and entanglement to compress features and associate semantics across heterogeneous data. 
The model transfer module employs quantum reinforcement learning to model decision-making and improve adaptability in dynamic environments. The federated aggregation module integrates quantum tensor decomposition with backpropagation-based corrections to provide privacy preservation with low overhead and to strengthen global model robustness. This work outlines a new paradigm for communication and model collaboration in future 6G intelligent transportation.
\end{abstract}


\section{Introduction}
With the rapid advancement of sixth-generation (6G) mobile communication technology, the Internet of Vehicles (IOV) is evolving from traditional single-vehicle connectivity to a comprehensive infrastructure that integrates vehicle-road-cloud-edge collaboration.
Through vehicle-to-everything (V2X) technology, vehicles, roadside units, edge servers, and cloud platforms achieve information sharing and collaborative operations, supporting key applications such as autonomous driving, collaborative decision-making, and road digital twins. Unlike traditional in-vehicle communication, IoV emphasizes semantic-level information exchange and collaborative optimisation between endpoints, edge, and cloud, aiming to enhance operational efficiency within limited bandwidth and latency constraints. As 6G development progresses, building an efficient, robust, scalable, and secure IoV system has become a key objective for advancing smart transportation.

However, in V2X networks, perception data sources are diverse, including camera video, lidar, millimeter-wave radar, and V2X messages. These data are high-dimensional, redundant, and heterogeneous, and direct transmission would significantly increase communication overhead \cite{r1}. Channel conditions are also highly uncertain due to factors such as geographical location, interference, and physical obstructions. Traditional bit-centric communication models can no longer satisfy V2X requirements for low latency, high reliability, and efficient bandwidth utilization. Meanwhile, vehicles and edge nodes must frequently exchange model parameters to support distributed collaborative perception and decision-making. However, in dynamic and heterogeneous networks, communication overhead from model aggregation, semantic inconsistencies across devices, and limited adaptability of model transfer severely restrict system intelligence and scalability.

To improve communication efficiency and coordination in V2X, researchers have increasingly adopted machine learning to optimize communication strategies, resource scheduling, and model updates. Although such methods enhance adaptability to some extent, they remain inadequate in highly dynamic and heterogeneous environments due to weak capabilities in processing high-dimensional data, slow training and inference, and poor generalization. Recently, quantum machine learning \cite{r2} has emerged as a promising paradigm that integrates quantum computing with machine learning. Leveraging quantum parallelism, superposition, and probabilistic properties, it offers unique advantages in processing high-dimensional data, adapting to low-resource scenarios, and optimizing complex models. Building on this, we propose a framework for intelligent communication and model co-optimization in V2X networks, aiming to construct an efficient, robust, and adaptive intelligent transportation system for 6G.

The proposed framework includes four key modules. (1) A channel-adaptive semantic communication algorithm based on quantum convolutional neural networks, which supports quantum AI-assisted semantic communication, enhances channel coupling, improves global generalization during data reconstruction, and provides effective semantic distortion measurement. (2) A quantum attention mechanism for multimodal data fusion, which aligns semantic information within and across modalities and performs cross-attention to compress and fuse heterogeneous data (e.g., camera and radar), thereby supplying richer input for collaborative perception and autonomous driving. (3) A model transfer and domain adaptation mechanism based on quantum Actor-Critic (AC). Here, model transfer is formulated as a quantum Markov decision process, enabling vehicles to determine transfer timing and methods autonomously. In contrast, a quantum domain alignment loss function ensures semantic space consistency across nodes and improves adaptability in dynamic environments.
(4) A model aggregation mechanism that integrates quantum tensor decomposition with secure communication. Within a federated learning (FL) framework, this design achieves low-overhead, high-efficiency multi-node collaboration and aggregation, while quantum tensor compression and parameter backpropagation ensure privacy and robustness of the global model.

The remainder of this paper is organized as follows: 
Section II reviews the latest research progress in semantic communication, multimodal fusion, model transfer, federated learning, and quantum machine learning in vehicle-to-everything communication; Sections III to VI detail the system architecture and design principles of the core modules proposed in this paper; Section VII presents experimental validation of quantum machine learning; and Section VIII summarizes the core contributions of this paper.

\begin{figure*}[htbp] 
    \centering
    \includegraphics[width=1\textwidth]{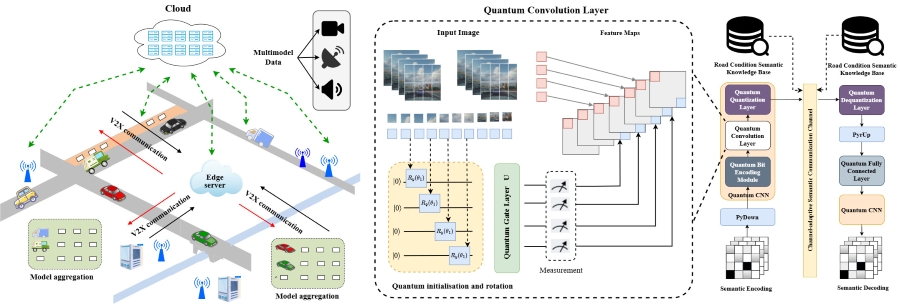} 
    \caption{Overall framework diagram of channel-adaptive quantum semantic communication algorithm.}
    \label{fig1}
\end{figure*}

\section{related work}

\subsection{Semantic Communication and Vehicle-to-Everything Communication Optimization}
Combining AI algorithms based on deep learning is a current research hotspot in semantic communication. Compared to traditional source-channel separation communication coding schemes, source-channel joint coding theoretically has the potential to achieve optimal solutions. Existing research \cite{r3} has attempted to jointly design source coding and channel coding, achieving systematic joint optimization of the coding process. Existing methods \cite{r4,r5} primarily rely on convolutional or self-attention structures to achieve semantic compression and robust transmission. However, in highly dynamic channel and heterogeneous device scenarios, these methods are prone to global semantic distortion, insufficient generalization to unseen environments, and limited adaptive capabilities under rapidly changing links.

\subsection{Multimodal Fusion and Cross-Modal Alignment}
Cross-modal fusion researches~\cite{r6,r7} typically achieve complementarity between modalities through attention, spatial alignment, and geometric priors. 
However, existing methods either prioritize real-time performance at the expense of reversibility and traceability or emphasize high-precision geometric alignment while struggling to meet the latency and computational constraints of in-vehicle systems. Section 5 of this paper proposes intra-modal self-attention based on quantum reversible modules, utilizing quantum entanglement to achieve cross-modal geometric alignment and missing data compensation. 
Additionally, quantum reversible cross-attention establishes a reversible mapping between decision-ready fusion semantics and reconstructible original data, thereby balancing real-time decision-making needs with accident reconstruction requirements.This bidirectional capability goes beyond the scope of most traditional designs that only perform forward fusion.

\subsection{Model Transfer, Domain Adaptation, and Reinforcement Learning}
Connected vehicle devices typically have very limited computing resources and are unable to quickly complete the calculations required for model fine-tuning locally. At the same time, cross-modal models have personalized requirements that cannot be met through centralized one-time training. Current research \cite{r8} attempts to model this task as a distributed game and uses game theory analysis methods to have different devices compete for resources to solve for the optimal strategy. Other studies~\cite{r9,r10} have attempted to introduce deep reinforcement learning, formulating the problem as a Markov decision process, and using multi-agent deep reinforcement learning to simulate each device connected to the network as an agent while training their strategies. Reinforcement learning is used for scheduling and strategy optimization, but in large-scale, multi-constrained vehicle-to-everything networks, training costs and convergence efficiency become major challenges. 

\subsection{Federated Learning and Efficient Aggregation}
Federated learning is widely used in vehicle-road-cloud collaboration \cite{r11}, but under conditions of limited edge computing power, upload constraints due to bandwidth/latency, and non-independent and identically distributed (Non-IID) data distribution, communication overhead and aggregation convergence stability remain key challenges. Existing compression and low-rank decomposition schemes \cite{r12,r13} can reduce transmission overhead, but common gradient/parameter pruning and quantization schemes may cause irreversible information loss and have limited adaptability to heterogeneous models. 

\subsection{Applications of Quantum Machine Learning in Communication and Collaboration}
Quantum machine learning \cite{r14,r15} has been introduced in recent years to tasks such as high-dimensional feature extraction, combinatorial optimization, and reinforcement learning acceleration. However, its systematic implementation in end-to-end links in the Internet of Vehicles remains relatively scarce.Existing explorations have primarily focused on individual capabilities (e.g., classification/regression using variational quantum circuits, strategy optimization using quantum reinforcement learning in small-scale environments), with insufficient attention given to the integrated implementation of the entire chain, including semantic communication, multimodal fusion, transfer alignment, and federated aggregation.

\section{Channel-adaptive quantum semantic communication algorithm}
To address the issues of channel dynamic changes and weak generalization in vehicle-to-everything semantic communication, this paper proposes to adopt real-time estimation of semantic channel states, a quantum semantic coding and decoding model, and quantum relative entropy to achieve quantum AI-assisted V2X semantic communication and channel coupling. Therefore, targeting the semantic communication requirements in V2X scenarios, the project first designs an adaptive semantic channel tailored for V2X and its optimization objectives. Based on the dynamic channel state of vehicles, it optimizes the semantic channel capacity caused by different communication protocols. Subsequently, it designs a V2X single-modal semantic encoding/decoding model and a semantic distortion measurement method based on a quantum convolutional neural network, achieving globally generalized semantic communication, as shown in Fig. \ref{fig1}.
\subsection{Adaptive Semantic Channels for Vehicle-to-Everything Communication and Their Optimisation Objectives}
To address the issues of semantic channels and channel capacity in vehicle-to-everything communication, we propose utilizing a traffic condition semantic knowledge base jointly maintained by V2X nodes to avoid the transmission of redundant information. This knowledge base helps reduce unnecessary data transmission in dynamic channels, thereby improving communication efficiency. To ensure that encoded semantic information can be effectively transmitted, we also need to achieve effective coupling with the physical channels of the V2X network. This process involves source compression based on shared traffic condition information and a semantic transmission method adapted to the V2X protocol.

Within this framework, improvements in semantic channel capacity depend not only on how information is encoded and transmitted, but also on real-time monitoring of uncertainty and signal quality during the transmission process. By dynamically adjusting communication protocols and encoding strategies, we can optimize signal transmission efficiency and ensure the reliable transmission of critical information. Additionally, the introduction of quantum technology offers a new approach to optimizing channel capacity. By leveraging quantum computing and reinforcement learning, we can measure and optimize semantic information transmission strategies in real time, further enhancing the communication capabilities of vehicle-to-everything networks.

\subsection{Quantum CNN Embedded Vehicle-to-Everything Single-Modality Semantic Coding and Decoding}
After semantically enhancing and optimizing the physical channel of the vehicle-to-everything network, the resulting semantic channel combines state estimation and adaptive adjustment with the characteristics of the semantic channel and the V2X physical channel. This section addresses the issue of local distortion in single-modal semantic encoding and decoding for vehicle-to-everything communication. 
It proposes introducing a quantum module into the convolutional neural network structure to enable the model to capture the key semantic features of vehicle perception data accurately. 
Consequently, we propose a globally generalized semantic communication algorithm leveraging quantum CNN embedding.


The quantum CNN network architecture designed in this paper is as follows: input data first passes through a quantum bit encoding module, which maps classical image pixel information to quantum states using angle encoding strategies; then enters the quantum convolution layer, where quantum gate operations composed of parameterized rotation gates and entangling CNOT gates are used to extract traffic condition semantic features, leveraging quantum entanglement properties to enable parallel extraction of high-dimensional features; subsequently, the quantum quantization layer converts the quantum state semantic features into classical signals via Pauli-Z expectation value measurements, which are transmitted via the V2X channel; at the receiving end, after decoding, the semantic features are input into the quantum neural network, where a quantum inverse transform is applied to maximize the fit with the original data and generate reconstructed data.

The advantages of quantum CNN are as follows: by leveraging the parallel computing capabilities of quantum computing, they enable faster feature extraction than traditional CNN, even with the limited computing power of edge devices in vehicle-to-everything networks. The superposition of quantum states makes the model more robust to changes in input data size, allowing it to adapt to differences in perception devices across different vehicle models.

\subsection{Distortion Measurement of Global Generalized Semantic Communication}
When measuring semantic distortion in vehicle-to-vehicle communication, it is important to focus on key semantics that affect driving safety. Traditional distortion indicators are unable to reflect semantic quality, so this project proposes to design a semantic information distortion measurement method based on quantum relative entropy. First, the quantum information entropy of the source data and the reconstructed data is calculated. To further measure semantic quality, the quantum information entropy is converted into quantum relative entropy.

This metric can effectively quantify the degree of distortion of key semantics. Subsequently, quantum relative entropy is used as a parameter to optimize the loss function of the encoding and decoding algorithm, improving the global generalization of semantic communication performance in vehicle-to-everything channel adaptation, and ensuring the reliability and security of semantic transmission in high-speed mobile scenarios.

\begin{figure*}[htbp] 
    \centering
    \includegraphics[width=1\textwidth]{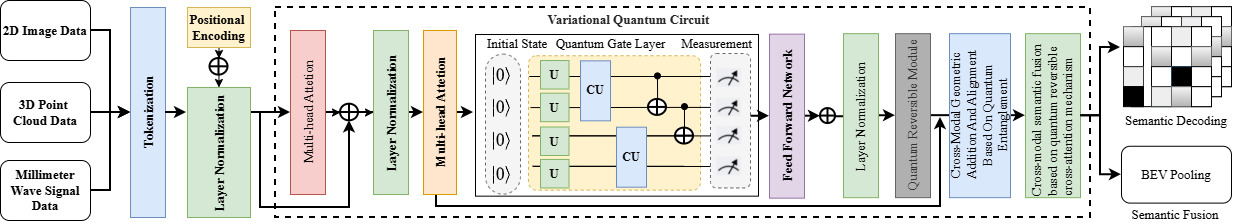} 
  
    \caption{Overall framework diagram of the joint source-channel coding and decoding model for cross-modal semantic fusion.}
    \label{fig2}
\end{figure*}

\section{Joint source-channel coding model with cross-modal semantic fusion}
To achieve efficient cross-modal semantic communication between heterogeneous devices in vehicle-to-everything networks, this paper proposes to design a joint source-channel coding and decoding model based on quantum technology for cross-modal semantic fusion. The focus is on addressing the geometric alignment and semantic fusion of multimodal data such as 2D images, 3D point clouds, and millimeter-wave signals, while simultaneously meeting the dual requirements of real-time traffic decision-making and data retrospection in V2X networks, as shown in Fig. \ref{fig2}.


\subsection{Modal Self-Attention Mechanism Based on Quantum Reversible Modules}

In vehicle-to-everything communication, different types of data must first undergo key feature extraction within each modality: 2D images from cameras must retain low-frequency semantic information such as road markings and pedestrian outlines; 3D point clouds from lidar must capture high-frequency details such as obstacle depth and vehicle dimensions; and millimeter-wave radar must focus on dynamic features such as distance and speed. This paper proposes to replace traditional feature mapping with quantum superposition state encoding, converting raw data from each modality into quantum states. By leveraging the superposition properties of quantum bits, both low-frequency core semantic information and high-frequency detail information can be stored simultaneously, enabling parallel extraction of multi-scale features without the need for complex filtering.

In the design of the modal attention mechanism, a quantum reversible module is introduced: utilizing the reversible evolution characteristics of quantum states, the original data's quantum state mapping relationship is retained during feature extraction, ensuring that complete features can be restored through quantum inverse operations during subsequent decoding. At the same time, a quantum dynamic sparse attention mechanism is adopted, which uses quantum measurement to prioritize key semantics, reducing redundant information transmission and adapting to the limited computing power of vehicle-to-everything edge devices.

\subsection{Cross-Modal Geometric Alignment Based on Quantum Entanglement}
In vehicle-to-everything communication, the spatial misalignment between 2D images and 3D point clouds is a key challenge in data fusion. This paper proposes to utilize quantum entanglement properties to achieve cross-modal geometric alignment modeling: by entangling the quantum states of the camera's 2D coordinates and the lidar's 3D coordinates, when the coordinates of either modality change due to vehicle movement, the entangled quantum state of the other modality will synchronously update, thereby achieving dynamic geometric alignment.

On this basis, quantum geometric patching technology is used to fill in the spatial intersection: for overlapping areas between 2D and 3D spaces, quantum superposition states are used to integrate the characteristics of both sides; for non-overlapping areas, quantum phase compensation technology is used to predict missing characteristics, ensuring the integrity of cross-modal geometric alignment and laying the spatial foundation for subsequent semantic fusion.


\subsection{Cross-Modal Semantic Fusion Based on a Quantum Reversible Cross-Attention Mechanism}
To address the requirement for semantic fusion that balances real-time decision-making and data restoration in vehicle-to-everything communication, we design a quantum reversible cross-attention mechanism to achieve cross-modal semantic fusion, where the cross-attention weights between modalities are generated through quantum phase modulation. 
For example, the obstacle type semantic information from the LiDAR 3D point cloud is encoded as a quantum phase, which is applied to the quantum state of the camera's 2D image. Through quantum interference effects, the semantic association between the two is enhanced.

This mechanism leverages the reversibility of quantum states to directly output fused semantics for real-time decision-making without decoding the original data. It can also fully restore the original multimodal data through quantum state inversion, meeting the data integrity requirements of scenarios such as accident reconstruction and liability determination. At the same time, the parallel computing characteristics of quantum computing significantly reduce the computational latency of cross-attention, ensuring real-time fusion efficiency in high-frequency dynamic scenarios in vehicle-to-everything networks.

Through the collaborative design of quantum reversible modules, entanglement alignment, and cross-attention, this solution can achieve efficient semantic fusion of cross-modal data in vehicle-to-everything networks, improving the accuracy and reliability of multi-device collaborative perception, and providing support for core scenarios such as vehicle-road collaboration and autonomous driving safety.

\begin{figure*}[htbp] 
    \centering
    \includegraphics[width=1\textwidth]{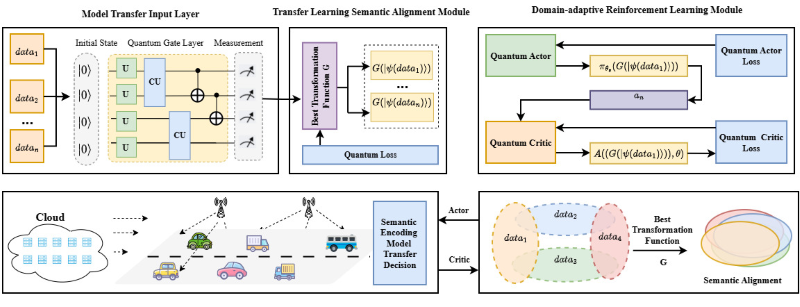} 
    \caption{Overall framework diagram of a cross-modal adaptive model transfer quantum reinforcement learning algorithm.}
    \label{fig3}
\end{figure*}

\section{Cross-modal domain adaptive model transfer based on quantum reinforcement learning}
In the context of vehicle-to-everything communication, vehicle nodes need to achieve real-time perception and collaborative decision-making through multimodal data. However, different vehicles exhibit significant domain differences due to varying road conditions, weather conditions, and traffic density, leading to semantic alignment discrepancies and insufficient adaptability when transferring semantic encoding models. To address the challenges of cross-modal model transfer under the high-dynamic, low-latency requirements of vehicle-to-everything networks, the project proposes to adopt domain-adaptive Markov decision process modeling combined with quantum Actor-Critic algorithms to achieve efficient model transfer and semantic alignment. The specific process is illustrated in Fig.~\ref{fig3}.

\subsection{Modelling of Vehicle-to-Vehicle Communication Model Transfer Issues Based on Markov Decision Processes}

In vehicle-to-everything networks, the transfer of semantic encoding models must be adapted to the high mobility of vehicle nodes and dynamic network topologies. The model transfer problem is modelled as a discrete-time Markov decision process, with the following core elements:

Actions: 
Define a set of model transition actions for a vehicle at a given moment, including local model fine-tuning, receiving global model updates from roadside units, and sharing local model parameters with neighboring vehicles. Action selection is driven by reward feedback from the quantum AC algorithm.

Reward: Considering the low-latency and high-reliability requirements of V2X networks, the reward function prioritises communication latency, decision accuracy, and data semantic consistency after model transfer, ensuring that rewards are strongly aligned with V2X safety and collaboration objectives.

State: The state space of a vehicle includes local multimodal data distribution characteristics, current channel quality, neighbouring node topology relationships, and local model performance metrics.

Optimisation objective: Minimise the overall cost of model migration, balancing latency and accuracy losses, and adapting to the real-time prioritisation characteristics of vehicle-to-everything networks.

\subsection{Cross-Modal Domain Adaptation and Semantic Alignment in Vehicle-to-Everything Communication Based on Transfer Learning}

In vehicle-to-everything networks, models often struggle to adapt to different data sources due to the disparate distribution of multimodal data. To address this issue, we propose a cross-modal adaptive method based on transfer learning, combining the quantum optimal transformation function $G(.)$ and the quantum universal loss function to achieve semantic alignment. Through $G(.)$, vehicle local data can be mapped to the quantum feature space, enhancing the accuracy of semantic matching and reducing information loss. The quantum universal loss function dynamically adjusts the weights of each modality during optimization to ensure optimal alignment during data fusion. The quantum optimal transformation function $G$ and the universal loss function work together to adapt in real-time to the dynamic environment of vehicle-to-everything networks, optimizing the fusion process of cross-modal data and ensuring that V2X nodes can efficiently adapt to different data distributions.

\begin{figure*}[htbp] 
    \centering
    \includegraphics[width=1\textwidth]{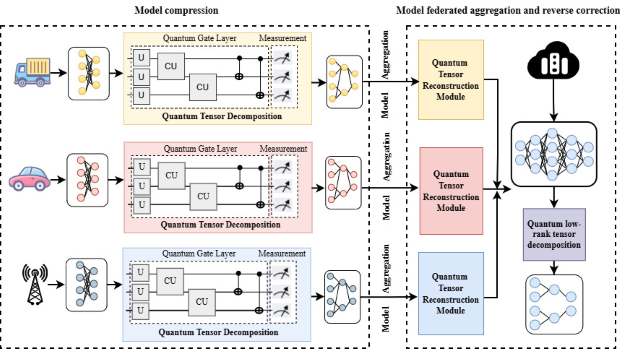} 
    \caption{Overall framework diagram of the quantum low-loss federated aggregation and reverse correction scheme for lightweight semantic encoding models.}
    \label{fig4}
\end{figure*}

\subsection{Solving Vehicle-to-Everything Model Transfer Problems Using Quantum Actor-Critic Algorithms}

To address the high-dimensional state space and real-time decision-making requirements of vehicle-to-everything networks, the quantum Actor-Critic algorithm is employed to replace traditional Actor-Critic algorithms. By leveraging the parallelism of quantum computing, the optimization efficiency of model transfer strategies is enhanced:

\textbf{Quantum Actor Network}: A quantum neural network is used to construct the policy network, with inputs encoded as quantum states representing vehicle states and outputs as quantum superposition states representing action probabilities. Quantum gate operations are used for feature extraction and action mapping, leveraging quantum superposition properties to parallel-evaluate the values of multiple actions, thereby accelerating strategy updates.

\textbf{Quantum Critic Network}: A value network is designed using a quantum neural network, with inputs being state-action quantum states and outputs being quantised state values. Quantum phase estimation techniques are employed to enhance value estimation accuracy, and quantum entanglement is used to capture non-linear correlations between states and actions.

\textbf{Quantum Gradient Update}: Optimises network parameters using a quantum gradient descent algorithm, obtains the expected values of policy gradients and value gradients through quantum measurement, and accelerates gradient solution through quantum parallel computing to meet the low-latency requirements of vehicle-to-everything communication.

In summary, by modelling the Markov decision process in the vehicle-to-everything scenario, designing a quantum semantic alignment module, and combining it with the efficient policy optimisation of the quantum Actor-Critic algorithm, we can achieve cross-modal domain adaptive model transfer, enhancing the adaptability and real-time performance of semantic encoding models in dynamic V2X environments.

\section{Quantum Low-Overhead Federated Aggregation and Backward Correction Algorithm Solution for Lightweight Semantic Encoding Models}

To address the issues of low aggregation efficiency and poor robustness in semantic encoding models caused by limited computing power and frequent communication among heterogeneous devices in vehicle-to-everything networks, the project proposes to design a quantum-based lightweight semantic encoding model, federated aggregation, and reverse correction scheme using quantum technology. By leveraging quantum lightweight modeling, quantum parameter updates, and quantum federated aggregation, the project aims to achieve efficient compression of distributed models, centralized aggregation, and global parameter correction in vehicle-to-everything networks, thereby meeting the high-dynamic, low-latency requirements of vehicle-to-infrastructure collaboration. The entire process is illustrated in Fig.~\ref{fig4}.

\subsection{Lightweight Modelling of a Vehicle-to-Everything Semantic Coding Model Based on Quantum Tensor Decomposition}

In vehicle-to-everything networks, edge devices, such as in-vehicle terminals and roadside units, have limited computing power and are unable to support complex semantic encoding models. The project utilises quantum tensor decomposition technology to perform lightweight modelling of semantic encoding models: model weights are mapped to quantum states, and high-dimensional weights are represented in low-rank quantum form through quantum superposition properties. By leveraging the parallelism of quantum bits, the model retains its core semantic encoding capabilities while significantly reducing the scale of model parameters, thereby adapting to the resource constraints of edge devices in vehicle-to-everything networks.

During the modelling process, the entanglement properties of quantum states are used to correlate parameters across model layers, ensuring that the compressed model can still accurately encode key semantics. Additionally, a quantum rank adaptation mechanism is introduced to dynamically adjust the rank reduction of quantum states based on real-time computational capabilities, balancing compression rate and encoding accuracy to meet the diverse requirements of different devices in the V2X environment.

\subsection{Semantic Encoding Parameter Update Based on the Quantum Dual Model Decomposition}

To achieve efficient local model iteration for vehicle-to-everything edge devices, the project designs a quantum dual-model decomposition mechanism: model parameters are divided into quantum core states and quantum auxiliary states. During local updates, edge devices perform lightweight iteration on core states through quantum state evolution while utilising quantum entanglement correlations in auxiliary states to reduce parameter transmission volume.

Specifically, devices such as in-vehicle terminals and roadside units collect road condition data locally, then optimise core state parameters via quantum gradient descent while simultaneously extracting correction information from auxiliary states through quantum measurement. This dual decomposition approach ensures rapid adaptation of local models to real-time road conditions while reducing communication overhead for parameter updates, thereby meeting the low-latency requirements of high-dynamic scenarios in vehicle-to-everything networks.

\subsection{Quantum Low-Latency Federal Aggregation and Reverse Correction of Semantic Coding Models for Vehicle-to-Everything Communication}

To address the requirement for global coordination of distributed device models in vehicle-to-everything networks, we develop a quantum low-rank federated aggregation framework: cloud servers act as aggregation centers, while edge devices serve as clients. In the initial phase, the cloud disseminates a global initial low-rank model via quantum broadcasting. In each aggregation round, clients utilise quantum double decomposition to compress local models and upload the low-rank quantum states to the cloud.

The cloud fuses the quantum states from all clients using quantum-secure aggregation technology to generate a globally optimised model, which is then reconstructed into a deployable semantic encoding model via quantum tensor reconstruction. To ensure alignment between edge devices and the global model, the cloud compresses the aggregated model using quantum low-rank decomposition and broadcasts it to all clients. Clients use a quantum reverse correction mechanism to fuse global model parameters with local models, achieving a closed-loop cloud aggregation and edge correction process to enhance the global consistency and robustness of vehicle-to-everything semantic encoding models.

Through the synergy of quantum lightweight modelling, dual decomposition parameter updates, and quantum federated aggregation, this solution reduces the parameter count and aggregation latency of vehicle-to-everything semantic encoding models, enabling low-latency, elastic, and scalable V2X semantic communication. 
The proposed framework provides efficient model support for multi-device collaborative perception and vehicle-road collaborative decision-making.

\begin{figure}[htbp] 
    \centering
    \includegraphics[width=0.5\textwidth]{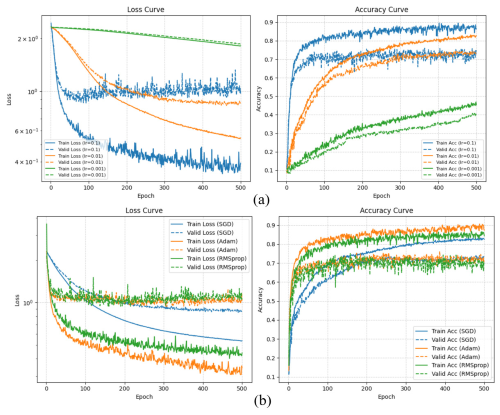} 
    \caption{The impact of different hyperparameters on model performances. (a) Learning rate experiments showing the convergence of loss and accuracy. (b) Optimizer experiments comparing Adam and other baseline optimizers.}
    \label{fig5}
\end{figure}

\section{experiment}

To validate the adaptability of quantum-enhanced models in the vehicle-to-everything optimization framework that integrates quantum machine learning, we conducted a controlled variable experiment on three core hyperparameters—batch size, learning rate, and optimizer—over 500 training cycles. By analyzing the dynamic changes in the loss curve and accuracy curve, we examined the impact of hyperparameters on model convergence and generalization, providing parameter configuration guidelines for model deployment in high-dynamic V2X scenarios.

Experimental setup: The classic handwritten digit dataset Digits was used to simulate lightweight perception data for vehicle-to-everything networks, which was split into training, validation, and test sets in a 6:2:2 ratio. The hardware setup consists of an Intel Core i9-13900K processor paired with an NVIDIA RTX 4090 graphics card. The software stack includes Python 3.10, PyTorch 2.8.8, and DeepQuantum 4.3. 
A quantum-classical hybrid convolutional neural network was constructed, featuring a 4-qubit feature extraction layer implemented as a variational quantum circuit within DeepQuantum and a classical post-processing layer connected through a differentiable interface to support end-to-end backpropagation. The core metrics are cross-entropy loss and accuracy rate.

Fig. 5(a) Learning rate experiments show that a learning rate of 0.01 results in synchronous decreases in loss and a stable accuracy of 0.8. A moderate step size avoids gradient issues and is suitable for adjusting quantum layer parameters. Fig. 5(b) Optimizer experiments indicate that the Adam optimizer performs optimally, with the lowest loss and a stable accuracy of 0.85. Its adaptive step size is suitable for the complex parameter space of quantum layers. 

Through three sets of hyperparameter optimization experiments, the optimal configuration for the quantum-enhanced model in this paper was determined to be: batch size 64, learning rate 0.01, and optimizer Adam. Under this configuration, the model achieved the lowest validation loss and highest accuracy, with a balanced convergence speed and generalization performance. The experimental results validate the effectiveness of quantum convolutional layers in lightweight models. Their parallel computing capabilities and quantum state superposition abilities enhance feature extraction efficiency. The optimal hyperparameter configuration provides a reliable model foundation for subsequent modules such as channel-adaptive semantic communication and multimodal fusion in vehicle-to-everything scenarios.

\section{Conclusion}
This paper addresses the key challenges in dynamic communication, multimodal fusion, and distributed collaboration in 6G vehicle-to-everything networks, and innovatively proposes a quantum machine learning fusion optimisation framework. By incorporating quantum parallel computing, entanglement properties, and reversible operation mechanisms, the framework systematically constructs four core modules: 1) A channel-adaptive semantic communication algorithm based on quantum convolutional neural networks, which utilises quantum relative entropy distortion metrics to enhance generalisation and channel capacity in high-mobility scenarios; 2) A quantum reversible attention-driven cross-modal fusion architecture that achieves efficient feature compression and semantic association of 2D/3D heterogeneous data through entanglement geometric alignment; 3) A quantum reinforcement transfer learning paradigm that models Markov decision processes using quantum Actor-Critic algorithms to enable rapid cross-domain adaptation in dynamic environments; 4) A lightweight federated aggregation supported by quantum tensor decomposition, combining dual model decomposition and reverse correction mechanisms to significantly reduce communication overhead and enhance model robustness. This framework theoretically breaks through the performance bottlenecks of Shannon information theory and classical machine learning, providing a highly reliable and low-latency quantum-enhanced solution for 6G intelligent transportation systems. In the future, it will focus on hardware embedding, noise suppression, and standardisation in the Noisy Intermediate-Scale Quantum (NISQ) era, driving quantum vehicle-to-everything communication from theoretical innovation to industrial implementation and empowering the digital transformation of intelligent transportation.

\section*{Acknowledgement}
\label{Acknowledgement}

This work was supported in part by the National Natural
Science Foundation of China (Nos. 62502101 and 62462002), and partially supported by the Natural Science Foundation of Guangxi, China (Nos. 2025GXNSFBA069394, 2025GXNSFAA069958).

\bibliographystyle{unsrt} \bibliography{reference}

\end{document}